\documentclass[twocolumn,prl,nofootinbib,superscriptaddress]{revtex4}

\usepackage{graphicx}
\usepackage{dcolumn}
\usepackage{bm}
\usepackage{epsfig}
\usepackage{amssymb}
\usepackage{amsmath}
\usepackage{subfigure}

\newcommand{\lsim}{\mathrel{\hbox{\rlap{\lower.55ex\hbox{$\sim$}} \kern-.3em \raise.4ex \hbox{$<$}}}}
\newcommand{\gsim}{\mathrel{\hbox{\rlap{\lower.55ex\hbox{$\sim$}} \kern-.3em \raise.4ex \hbox{$>$}}}}
\newcommand{\beq}{\begin{equation}}
\newcommand{\eeq}{\end{equation}}
\newcommand{\beqa}{\begin{eqnarray}}
\newcommand{\eeqa}{\end{eqnarray}}

\newcommand{\Mpl}{M_\mathrm{Pl}}

\newcommand{\gtil}{\tilde{g}}

\newcommand{\initT}{T_{J,i}}
\newcommand{\phimin}{\phi_\mathrm{min}}
\newcommand{\omk}{\omega_k}
\newcommand{\ak}{\alpha_k}

\newcommand{\phibar}{\bar{\phi}}
\newcommand{\kp}{k_\mathrm{pk}}
\newcommand{\phib}{\phi_\mathrm{ta}}
\newcommand{\vel}{\dot{\phi}_M}

\begin{document}

\title{Catastrophic Consequences of Kicking the Chameleon} 

\author{Adrienne L. Erickcek}\email{erickcek@cita.utoronto.ca}
\affiliation{CITA, University of Toronto, 60 St.~George Street, Toronto, Ontario M5S 3H8, Canada}
\affiliation{Perimeter Institute for Theoretical Physics, 31 Caroline St. N, Waterloo, Ontario N2L 2Y5, Canada}
\author{Neil Barnaby}\email{n.barnaby@damtp.cam.ac.uk}
\affiliation{DAMTP, Cambridge University, Wilberforce Road, Cambridge, CB3 0WA, United Kingdom
}
\author{Clare Burrage}\email{Clare.Burrage@nottingham.ac.uk}
\affiliation{School of Physics and Astronomy, University of Nottingham, Nottingham, NG7 2RD, United Kingdom}
\author{Zhiqi Huang}\email{zhiqi.huang@cea.fr}
\affiliation{Institut de Physique Theorique, CEA/Saclay, Orme des Merisiers, Gif-sur-Yvette, France}

\begin{abstract}
The physics of the ``dark energy'' that drives the current cosmological acceleration remains mysterious, and the dark sector may involve new light dynamical fields.  If these light scalars couple to matter, a screening mechanism must prevent them from mediating an unacceptably strong fifth force locally.  Here we consider a concrete example: the \emph{chameleon mechanism}.  We show that the \emph{same} coupling between the chameleon field and matter employed by the screening mechanism also has catastrophic consequences for the chameleon during the Universe's first minutes.  The chameleon couples to the trace of the stress-energy tensor, which is temporarily non-zero in a radiation-dominated Universe whenever a particle species becomes non-relativistic.  These ``kicks'' impart a significant velocity to the chameleon field, causing its effective mass to vary non-adiabatically and resulting in the copious production of quantum fluctuations.  Dissipative effects strongly modify the background evolution of the chameleon field, invalidating all previous classical treatments of chameleon cosmology.  Moreover, the resulting fluctuations have extremely high characteristic energies, which casts serious doubt on the validity of the effective theory.  
Our results demonstrate that quantum particle production can profoundly affect scalar-tensor gravity, a possibility not previously considered.
Working in this new context, we also develop the theory and numerics of particle production in the regime of strong dissipation.
\end{abstract}

\maketitle
 
{\bf Introduction--}  
Understanding cosmic acceleration \cite{Riess:1998cb, Perlmutter:1998np} is one of the deepest open problems in cosmology.
Several theories postulate that the dark energy responsible for this acceleration is sourced by a dynamical scalar field \cite{Wetterich:1994bg, Zlatev:1998tr, Amendola:1999er, Caldwell:2009ix}.  
Such theories face a severe challenge: a light scalar field will generally mediate a long-range fifth force that is subject to stringent experimental bounds \cite{Adelberger:2009zz}.  If the scalar field couples to matter, then the theory must include a ``screening mechanism" that prevents the scalar field from mediating a long-range force in local environments \cite{Jain:2010ka}.

Here we consider the \emph{chameleon} \cite{Khoury:2003aq,Khoury:2003rn}, a well-studied screening mechanism that is essential to $f(R)$ gravity \cite{Carroll:2003wy,Chiba:2006jp,Faulkner:2006ub,Hu:2007nk, Brax:2008hh}. 
We show that the \emph{same} coupling between the chameleon scalar field and matter that enables the screening mechanism nearly always leads to a 
breakdown of calculability just prior to Big Bang Nucleosynthesis (BBN).  If this coupling is not too weak, quantum fluctuations of the
chameleon field inevitably become excited when particles become non-relativistic.  Weakly coupled chameleons require finely-tuned initial conditions to 
avoid the same fate.  The produced fluctuations contain a significant fraction of the chameleon's energy, showing that the field cannot generically be 
treated as a homogeneous classical condensate, as was assumed in all previous works \cite{Brax:2004qh, Mota:2011nh}. 
Moreover, the characteristic momenta of fluctuations can \emph{exceed} the Planck scale for typical parameters, casting serious doubts on the validity 
of Effective Field Theory (EFT).  This trans-Planckian regime includes all strongly coupled chameleons \cite{Mota:2006ed, Mota:2006fz}, which are the models relevant for direct detection experiments \cite{Steffen:2010ze, Brax:2010xx, Rybka:2010ah, Baker:2012nq}.

In chameleon gravity, the spacetime metric $\gtil_{\mu\nu}$ that governs geodesic motion differs from the metric $g_{\mu\nu}$ in the Einstein-Hilbert 
action; $\gtil_{\mu\nu} = \exp[2\beta\phi/\Mpl] g_{\mu\nu}$, where $\phi$ is the chameleon field, $\beta$ is a dimensionless coupling, and 
$\Mpl$ is the Planck mass.  The Lagrangian is
\begin{equation}
\label{L}
  \mathcal{L} = \frac{\Mpl^2}{2}R[g_{\mu\nu}] -\frac{1}{2}(\partial\phi)^2 - V(\phi) 
  + \mathcal{L}_{\mathrm{mat}}\left[\tilde{g}_{\mu\nu}, \psi_m^{(i)} \right] \, .
\end{equation}
The chameleon potential $V(\phi)$ has to have a particular form for the screening mechanism to work successfully.  A typical example of this class of potentials is
\beq
V(\phi) = M^4 \exp\left[\left(M / \phi \right)^n\right] \, , \hspace{2mm} n > 0 \, .
\label{pot}
\eeq
We assume $\beta\geq \mathcal{O}(10^{-2})$ so that the screening mechanism is relevant.
Evading fifth-force constraints requires $M \lsim 0.01\,\mathrm{eV}$ \cite{Mota:2006fz}, and if $M \simeq0.001\, \mathrm{eV}$, the 
chameleon drives late-time cosmic acceleration \cite{Brax:2004qh}.  E\"{o}t-Wash experiments also constrain $n$ and $\beta$ \cite{Gannouji:2010fc, Upadhye:2012qu}.

In a Friedmann-Roberston-Walker spacetime ($g_{\mu\nu}dx^\mu dx^\nu = -dt^2 + a^2(t)d{\bf x}^2$),
\beq
\ddot{\phi}+3H\dot{\phi} - a^{-2} \nabla^2 \phi  = -\left[ V^\prime(\phi) + \beta (\rho-3P) / \Mpl \right] \, ,
\label{eom}
\eeq
where $\dot{\phi}\equiv \partial_t\phi$, $H \equiv \dot{a} / a$, and $\rho$ and $P$ are the ``Einstein-frame" energy density and pressure, which are related to the observed (Jordan-frame) energy density $\rho_J$ and pressure $P_J$ by $\rho/\rho_J = P/P_J = \exp[4\beta\phi/\Mpl]$.  The $\phi$ dynamics are governed by an effective potential, $V_{\mathrm{eff}} = V + \beta \phi (\rho-3P) / \Mpl$, whose minimum, $\phi_{\mathrm{min}}$, depends on $\rho$ and $P$.  The chameleon mass, $m_\phi^2 \equiv V_{\mathrm{eff}}''(\phi_{\mathrm{min}})$, increases with $\rho$.  In high density regions, $\phi$ is heavy and cannot mediate a long-range force \cite{Khoury:2003aq,Khoury:2003rn}.  

The potential $V(\phi)$ was designed to provide this screening mechanism and does not originate from fundamental physics.  While there have been attempts to realize Eqs.~(\ref{L},\ref{pot}) in string theory \cite{Brax:2006np,Hinterbichler:2010wu}, chameleon gravity is usually treated as a low-energy EFT.  Quantum effects were ignored until recently \cite{Upadhye:2012vh}.  We consider a very different kind of quantum effect, related to particle production in a time-dependent background.

{\bf Kicks--}  
We assume that the Universe became radiation dominated at a high temperature ($T\gsim \mathrm{TeV}$), and at this time, $\phi$ was a classical, homogeneous condensate with $M \ll \phi_i \lsim \Mpl$ \cite{Brax:2004qh}.  (If $\phi_i\ll M$, then the force $V'$ pushes $\phi$ to larger values.)  Prior to BBN, $\phimin \lsim M$, but Hubble friction prevents the chameleon from rolling toward $\phimin$ while $(\rho-3P) \ll \rho$.  This is problematic because $\phimin \ll \Mpl$ today, and variations in $\phi$ can be interpreted as variations in particle masses.  To avoid spoiling the success of BBN, the chameleon must reach $\phi \lsim 0.1 \Mpl/\beta$ before the temperature cools to a few MeV \cite{Brax:2004qh}.  Since $\phi_i$ is set by unknown physics in the very early Universe, some mechanism to displace $\phi$ prior to BBN is usually required to satisfy this constraint.

\begin{figure}
 \centering
 \resizebox{3.4in}{!}
 {
      \includegraphics{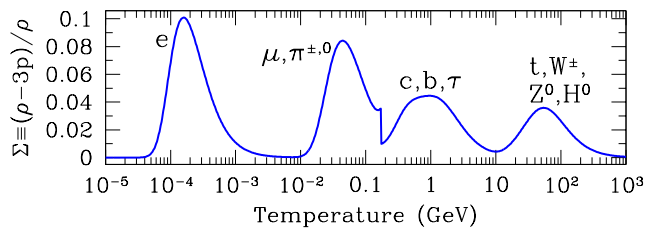}
 }
\caption{The kick function $\Sigma=(\rho-3P) / \rho$ vs Jordan-frame temperature.  We account for all SM particles. The discontinuity at $T_J = 170 \, \mathrm{MeV}$ corresponds to the QCD phase transition.}
\label{Fig:SMkicks}
\end{figure}

Fortunately, there is an effect that will ``kick'' $\phi$ to smaller values \cite{Damour:1993id,Damour:1992kf,Brax:2004qh}.  The quantity $\Sigma \equiv (\rho-3P)/\rho$ becomes temporarily non-zero when the radiation temperature drops below the mass of a species X in thermal equilibrium; at higher temperatures $\Sigma$ is small because $P_X\approx \rho_X/3$ and at lower temperatures it is small because $\rho_X$ is Boltzmann suppressed.  At this time, the last term in Eq.~(\ref{eom}) overcomes the Hubble friction, and $\phi$ rolls toward $\phimin$.  In Fig.~\ref{Fig:SMkicks} we include all Standard Model (SM) particles and plot $\Sigma$ as a function of the Jordan-frame temperature $T_J$; the contributions from individual particles merge into four distinct kicks \cite{Coc:2006rt, longdraft}.

We solved Eq.~(\ref{eom}) numerically for a homogeneous chameleon with a wide range of initial values.  We find that the kicks generically drive $\phi$ to $\phi\lsim M$, where $V(\phi)$ becomes important.  At this moment, the chameleon's velocity $\dot{\phi}$ is \emph{much} larger than $M^2$, the scale that controls $V(\phi)$.  This huge velocity causes the chameleon mass to vary rapidly, and then particle production spoils the classical approximation.

{\bf Chameleon Velocities--} 
Before discussing particle production, we must understand why the kicks drive $\phi$ toward the potential barrier at $\phi\lsim M$ with a large velocity.  
If  $\phi_i \gg \phimin$, then we can neglect the $V^\prime(\phi)$ term in Eq.~(\ref{eom}).   Since $V(\phi) \ll \rho$ and $\Sigma \ll 1$, the homogeneous dynamics are well approximated by
\beq
\varphi^{\prime\prime}+\varphi^\prime\left[1-(\varphi^\prime)^2 / 6\right]= -3\left[1-(\varphi^\prime)^2 /6\right]\beta\Sigma(T_J),
\label{redeom}
\eeq
where $\varphi \equiv \phi/\Mpl$, $\varphi'\equiv \partial_p \varphi$, and $p\equiv\ln(a/a_i)$.  The Jordan-frame temperature ($\rho_J\propto T_J^4$) also depends on $\varphi$: 
\beq
T_J = \left[\frac{g_{*S}(\initT)}{g_{*S}(T_{J})}\right]^{1/3} \initT e^{\beta(\varphi_i-\varphi)} e^{-p},
\label{TJ}
\eeq
where $g_{*S}(T_{J})$ is the entropy density divided by $(2\pi^2/45)T_J^3$, and $\initT$ and $\varphi_i$ are initial conditions.

These equations admit a novel \emph{surfing solution}, characterized by a constant Jordan-frame temperature:
\begin{equation}
\label{surfer}
  \varphi_s'(p) = -\beta^{-1} \hspace{2mm}\Leftrightarrow\hspace{2mm}T_s \equiv T_J\left[\varphi_s(p)\right] = \mathrm{const.}
\end{equation}
This ansatz solves Eq.~(\ref{redeom}) if $\Sigma(T_s) = 1/(3\beta^2)$.  If only the SM contributes to $\Sigma$, then surfing solutions exist for $\beta > 1.82$.  Numerically solving Eq.~(\ref{eom}) with $\beta>1.82$ confirms that the surfing solution is an attractor if \mbox{$\dot{\phi}^2\ll\rho$} prior to the kicks.  Previous studies \cite{Brax:2004qh, Mota:2011nh} missed the surfing solution because they neglected the $\varphi$ dependence in Eq.~(\ref{TJ}).

Chameleons with $\beta>1.82$ can ``surf'' the kick function from an arbitrarily large initial condition; $\varphi'(p) = -\beta^{-1}$ until $\varphi \simeq \phimin/\Mpl$, where $V^\prime$ becomes important and Eq.~(\ref{redeom}) breaks down.  If $\beta\geq3.07$, then $T_s > 61$ GeV and the chameleon quickly settles into the surfing solution.  If $1.82 <\beta < 3.06$, the chameleon will not surf the first kick, but it can surf a subsequent kick if earlier kicks leave $\phi\gg\phimin$.  Consequently, any chameleon with $\beta>1.82$ will reach $\phimin$, regardless of $\phi_i$.  

If the chameleon cannot surf, then the kicks displace $\phi$ by a finite amount \cite{longdraft}.  However, any chameleon with $\beta > 0.42$ will reach $\phimin$ during the last kick if $\varphi < (0.1/\beta)$ prior to that kick, as required by BBN.  Chameleons with $\beta<0.42$ can avoid colliding with the potential wall, but only if their initial condition is finely tuned so that all the kicks from particles with masses $>\!\!\mathrm{MeV}$ leave $0.56\beta < \varphi < 0.1/\beta$.  For $f(R)$ gravity, $\beta = 1/\sqrt{6}$, and impact can only be avoided if $0.23 < \varphi < 0.24$ prior to BBN.

Having established that the kicks almost always take the chameleon to $\phimin$, we now consider the chameleon's velocity when it impacts its bare potential: 
\mbox{$\dot{\phi}\approx -0.6 g_*^{1/2} \varphi' \left[ 3-0.5(\varphi')^2 \right]^{-1/2} \, T_J^2$}.  Typical velocities are controlled by $T_J$, evaluated when 
$\phi=\phi_{\mathrm{min}}$.  Quantitatively $T_J\gsim 0.5\, \mathrm{MeV}$ at this time and $|\varphi'| > 0.02$, unless a kick deposits $\phi$ exactly at 
$\phimin$, so $|\dot{\phi}|^{1/2} > 0.07 \, \mathrm{MeV} \gg M$ in all but a few finely-tuned cases. Moreover, $|\dot\phi|$ is usually much larger;  a surfing 
chameleon with $\beta\geq3.07$ has $|\dot{\phi}|^{1/2} > 63\, \mathrm{GeV}$ at impact.  

{\bf Particle Production--} When the chameleon reaches $\phimin$ with $|\dot{\phi}|\gg M^2$, it climbs up the steep side of its effective potential until its kinetic energy is exhausted, and then it rolls back to larger values.  This ``reflection'' occurs on a \emph{very} short time scale, so we can neglect the expansion of the Universe.  The production of quantum fluctuations $\delta\phi(t,\vec{x}) = \phi(t,\vec{x}) - \phibar(t)$ is governed by their effective mass: $m_\phi^2(t) \equiv V_\mathrm{eff}''\left[\phibar(t)\right] \simeq V''\left[\phibar(t)\right]$ when $\phi \lsim M$.  Near the moment of reflection, $m^2_\phi$ changes significantly over a tiny time scale $\Delta t \sim V'' / (V''' \vel)$, where $\vel$ is the chameleon's velocity when it starts its climb ($\phi \simeq M$).  Such non-adiabatic variation will excite modes with $k \lsim (\Delta t)^{-1} \sim |\vel| / M$ \cite{Kofman:1997yn}.  The perturbation energy per logarithmic interval in $k$ is $E_k = k^3 \omega_k n_k / (2\pi^2)$ where $n_k$ is the occupation number.  Since $|\vel| \gg M^2$, these modes carry a tremendous amount of energy; unless $n_k\ll1$, $E_k$ greatly exceeds the energy of the chameleon field prior to the reflection.   Therefore, we expect the rapid turn-around of $\phibar$ to generate fluctuations with \emph{very} high energies that strongly backreact on the background $\phibar$ even when their occupation numbers are tiny.

To make these heuristic claims quantitative, we express $\delta\phi$ in terms of creation and annihilation operators,
\beq
\delta\phi(t,\vec{x})= \int \frac{d^3k}{(2\pi)^{3}}\left[\hat{a}_{\vec{k}} \phi_k(t) e^{i\vec{k}\cdot\vec{r}} + \hat{a}^{\dagger}_{\vec{k}} \phi^*_k(t) e^{-i\vec{k}\cdot\vec{r}} \right].
\eeq
If we neglect non-linear $\delta \phi$ interactions while keeping the leading-order backreaction of $\delta\phi$ on $\phibar$, Eq.~(\ref{eom}) implies
\beqa
&&\ddot{\phibar} + V_{\mathrm{eff}}^\prime(\phibar) + \frac{1}{2} V^{\prime\prime\prime}_{\mathrm{eff}}(\phibar)\langle \delta\phi^2 \rangle = 0, \label{bkgrd}\\
&&\langle\delta\phi^2 \rangle =  \int_{k>k_{\mathrm{IR}}} \frac{d^3k}{(2\pi)^3}\left( |\phi_k|^2 - \frac{1}{2\omk}\right), \label{variance}\\
&&\ddot{\phi}_k+\omega_k^2\phi_k = 0 \, , \hspace{3mm}\omega_k(t)^2 \equiv k^2+V^{\prime\prime}_{\mathrm{eff}}\left[\phibar(t)\right]\label{lineq}.
\eeqa
Eq.~(\ref{bkgrd}) is the spatial average of the second-order Taylor expansion of Eq.~(\ref{eom}) around $\phi = \phibar$.  Including the $\langle \delta\phi^2 \rangle$ term ensures that production of fluctuations drains energy from $\phibar$: $\frac{d}{dt}\bar{\rho} = -\frac{d}{dt}\langle \delta \rho \rangle$.  Eq.~(\ref{variance}) has been regulated as in previous works \cite{Kofman:2004yc,Barnaby:2010ke}, and $k_{\mathrm{IR}}$ is the scale on which we coarse-grain the chameleon; modes with 
$k < k_{\mathrm{IR}}$ are absorbed into the background $\phibar(t)$.  Note that we omit mode-mode couplings for fluctuations with 
$k > k_{\mathrm{IR}}$, whereas the coarse-grained field obeys the nonlinear Eq.~(\ref{bkgrd}).  Eq.~(\ref{lineq}) is solved with vacuum initial conditions,
$\phi_k = e^{-i\int^t\omega_k(t')dt'} / \sqrt{2\omega_k(t)}$, prior to particle production.

We solved the closed system (\ref{bkgrd}-\ref{lineq}) numerically, allowing the classical trajectory $\phibar(t)$ to reflect off the potential barrier near 
$\phi=0$.  We take $V_{\mathrm{eff}}(\phi)$ to be given by Eq.~(\ref{pot}) with $M=0.001$ eV and $2\leq n \leq 10$; the matter coupling is irrelevant because we are only concerned with 
small field displacements ($\Delta\phi \lsim M$) \cite{longdraft}.  We start the evolution at $\phibar=2M$ and we take $\vel$ as determined by the kick 
dynamics discussed earlier.  We integrate modes with $k_{\mathrm{IR}} < k < k_{\mathrm{max}}$ where $k_{\mathrm{max}} \gg (\Delta t)^{-1}$ (the short 
time scale of the reflection).  We take $k_{\mathrm{IR}} < 0.05 (\Delta t)^{-1}$ to capture the evolution of the modes that are most copiously produced 
while minimizing the errors introduced by neglecting the mode-mode couplings.

\begin{figure}
 \centering
 \resizebox{3.4in}{!}
 {
      \includegraphics{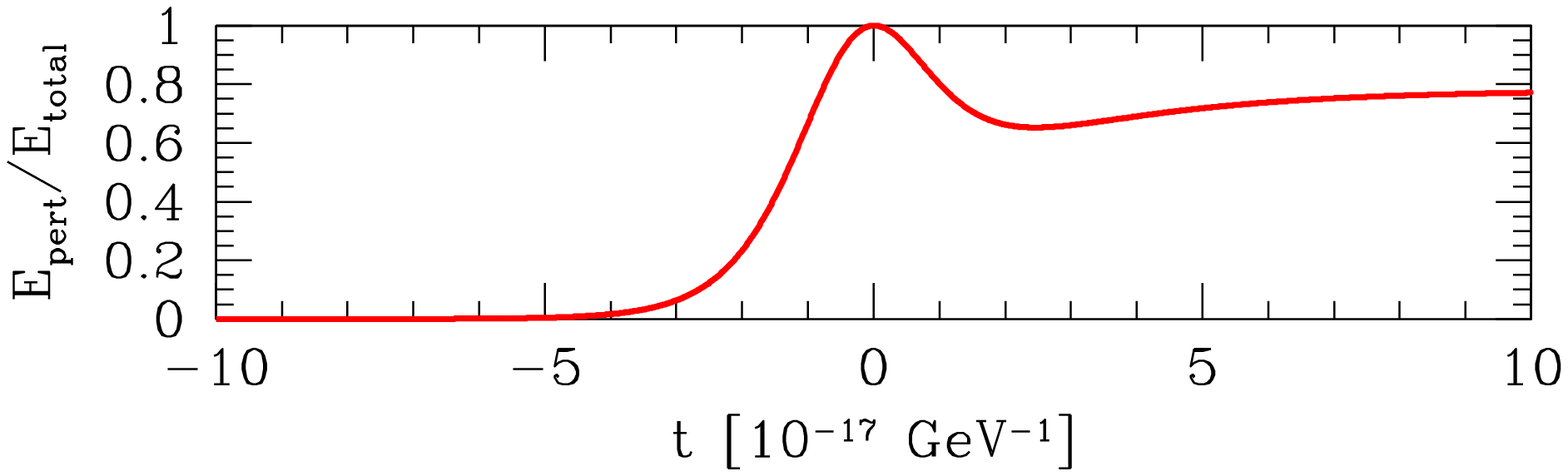}
 }
\caption{Time evolution of the total energy in fluctuations divided by the total energy in the chameleon field for $n=2$, $\vel=100\,\mathrm{GeV^2}$ and 
$k_{\mathrm{IR}}= 10^{15}$ GeV; $\phibar$ turns around when $t=0$.}
\label{Fig:Etot}
\end{figure}

Figure~\ref{Fig:Etot} shows the time evolution of the total energy in fluctuations ($E_{\mathrm{pert}}$); as $\phi$ climbs its potential, $E_{\mathrm{pert}}$ 
grows from zero to become an $\mathcal{O}(1)$ fraction of the background energy even before $\phibar$ turns around at $t=0$.  This energy transfer 
clearly indicates that one cannot treat $\phi$ as a homogeneous, classical field.  Shortly after the reflection, interactions become strong, and 
Eqs.~(\ref{bkgrd}-\ref{lineq}) break down.  Naively extrapolating our results into this (uncalculable) regime, we see that $E_{\mathrm{pert}}$ 
eventually reaches a steady state.  The asymptotic value of $E_{\mathrm{pert}}$ depends on $k_{\mathrm{IR}}$, precisely because nonlinearities are 
important 
and $k_{\mathrm{IR}}$ sets the longest wavelength mode that is treated linearly.  The choice of $k_{\mathrm{IR}}$ does \emph{not} affect our claim about 
the breakdown of 
the classical approximation, which happens before the reflection point.  Any attempt to follow the chameleon through its reflection off the potential wall 
must provide an account of particle production in the regime of strong dissipation and strong nonlinearity.

\begin{figure}
 \centering
 \resizebox{3.4in}{!}
 {
      \includegraphics{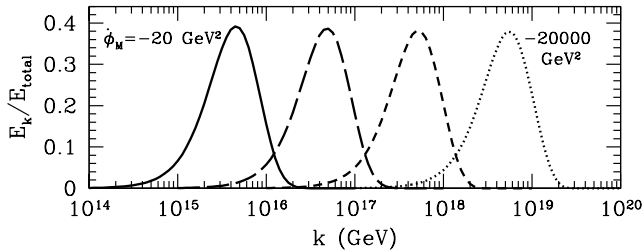}
 }
\caption{Energy spectra for initial chameleon velocities $\vel  /\mathrm{GeV}^2= -20$ (solid), -200 (long-dashed), -2000 (short-dashed), and -20000 (dotted).  The energy per logarithmic interval in $k$ ($E_k$) is shown as a fraction of the total chameleon energy $E_\mathrm{total} \simeq \vel^2/2$.  In all cases, $k_\mathrm{IR}/\kp = 0.02$ and $n=2$.}
\label{Fig:spectra}
\end{figure}

Figure~\ref{Fig:spectra} shows the energy spectra of produced particles for different values of $\vel$.  As expected, the perturbation energy spectrum peaks 
at a very high wavenumber ($\kp$) that depends on the chameleon's initial velocity.  Although $n_k\ll 1$, the energy in perturbations is substantial 
because their typical momenta are large; even for modest $\vel$, most of the energy in fluctuations is in modes with 
$k=\kp \gsim 10^{15}\mathrm{GeV}$!  Changing $k_{\mathrm{IR}}$ does not affect the shape of the spectra, which indicates that nonlinear 
interactions do not change which modes are excited.  Rather, $\kp$ is determined by the timing of the reflection.

{\bf Analytic Method--}   We now derive an analytic model of the reflection that gives an expression for $\kp$.  Writing $\phi_k$ using time dependent Bogoliubov coefficients gives
\beq
\label{bog}
\phi_k = \frac{\alpha_k(t)}{\sqrt{2\omk}}e^{-i\int^t \omk(t^\prime) dt^\prime}+\frac{\beta_k(t)}{\sqrt{2\omk}}e^{+i\int^t \omk(t^\prime) dt^\prime} \, ,
\eeq
where $|\alpha_k|^2 -|\beta_k|^2 =1$ and $n_k(t)$ equals $|\beta_k(t)|^2$.  Rapid changes in $\omk$ excite perturbations because \cite{Kofman:1997yn}
\beq
\dot{\beta}_k = \frac{\dot{\omega}_k}{2\omk} e^{-2i\int^t \omk(t^\prime) dt^\prime}\ak \, .
\label{betadot}
\eeq
Since $n_k \ll 1$ in our case, $|\beta_k| \ll |\alpha_k|$, and we may integrate Eq.~(\ref{betadot}) with $\alpha_k=1$ to obtain an approximation for $\beta_k(t)$ \cite{Braden:2010wd}.  Using this solution, along with (\ref{bog}), we compute $\langle\delta\phi^2 \rangle$ and write Eq.~(\ref{bkgrd}) in a closed form:
\beq
  \ddot{\phibar} + V_{\mathrm{eff}}^\prime(\phibar) 
  = V^{\prime\prime\prime}[\phibar(t)] \int_0^t V^{\prime\prime\prime}[\phibar(t^\prime)] \dot{\bar{\phi}}(t^\prime) K(t-t') dt^\prime, \nonumber
\eeq
where $K(x) = \mathrm{CosineIntegral}\left[{2 k_\mathrm{IR}x}\right] / (16\pi^2)$.  The right hand side represents dissipation from particle production and matches the 1-loop dissipation term computed in Ref.~\cite{Boyanovsky:1994me} using a different method.

\begin{figure}
 \centering
 \resizebox{3.4in}{!}
 {
      \includegraphics{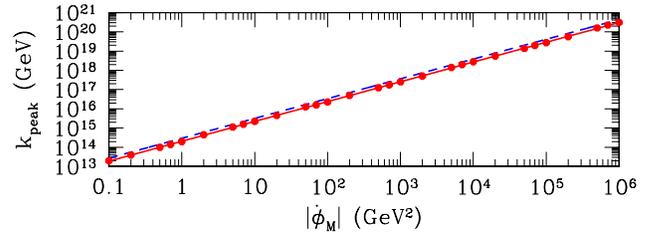}
 }
\caption{The peak wavenumber in the perturbation spectrum as a function of the chameleon initial velocity $\vel$ for $n=2$. The points show the numerical results.
The dashed line is Eq.~(\ref{kmodel}), and the solid line is Eq.~(\ref{kmodel}) multiplied by 0.7.}
\label{Fig:kpeakModel}
\end{figure}

The magnitude of the dissipation term increases sharply as $\phibar$ decreases, so we may restrict our analysis to a short time just before $\phibar$ 
turns around.  In that regime, we can integrate by parts to approximate Eq.~(\ref{bkgrd}) as 
$\ddot{\phibar} + V_{\mathrm{eff}}'(\phibar) + \kappa(t)V^{\prime\prime\prime}(\phibar)V^{\prime\prime}(\phibar) \approx 0$, where $\kappa(t)$ depends 
logarithmically on $k_\mathrm{IR}t$ with $0.02 \lsim \kappa \lsim 0.05$ \cite{longdraft}.  Neglecting the slow evolution of $\kappa$, particle production
effectively changes the chameleon potential to $V(\phi)+V_D(\phi)$ with $V_D(\phi)\simeq (\kappa/2)[V^{\prime\prime}(\phi)]^2$.   For $\phi \lsim M$, $V_D(\phi)\gg V(\phi)$; the chameleon dynamics are dominated by quantum effects.  Indeed, the numerical solutions confirm that the turn-around point $\phibar=\phib$ has $V_D(\phib) = \vel^2/2$ with $\kappa \simeq 0.03$.  Therefore, we should use $V_D(\phi)$ when computing $\kp \sim (\Delta t)^{-1} \approx \sqrt{V_D^{\prime\prime}(\phib)}$;
\beqa
\kp \simeq \frac{n|\vel|}{2M} \left[\frac{M}{\phib}\right]^{n+1} \simeq \frac{n b_n|\vel|}{2M} \ln^{\frac{n+1}n}\left[\frac{\vel^2}{n^4\kappa M^4}\right],
\label{kmodel}
\eeqa
where $b_n$ is an order-unity constant; this final approximation is accurate for \mbox{$10^{-6}<|\vel|/\mathrm{GeV}^2<10^6$}.
Figure~\ref{Fig:kpeakModel} shows that this analytic result successfully matches the numerics for $n=2$, up to a numerical factor that is close to unity.  This model is similarly successful for other values of $n\leq10$, and increasing $n$ changes $\kp$ by less than 25\% over the relevant $|\vel|$ range.

\begin{figure}
 \centering
 \resizebox{3.4in}{!}
 {
      \includegraphics{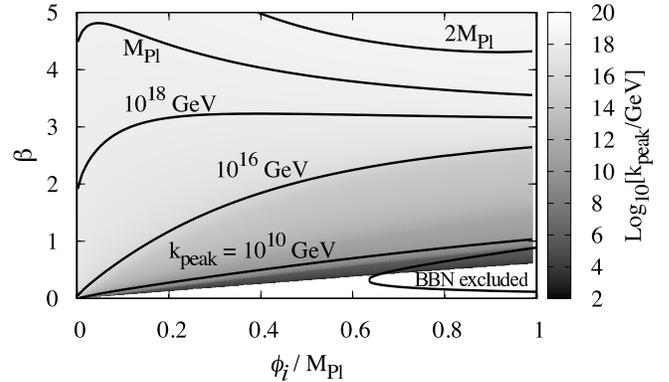}
 }
\caption{Peak wavenumber ($\kp$) in the perturbation energy spectrum, as a function of initial position ($\phi_i$) and coupling constant ($\beta$) for $n=2$.  
The white region shows values of $\phi_i$ sufficiently large that $\phi \gg M$ after all four kicks. The region marked ``BBN excluded" is forbidden 
because $\phi > 0.1 \Mpl/\beta$ prior to the last kick, which spoils the success of BBN.}
\label{Fig:kpeakContour}
\end{figure}

In Fig.~\ref{Fig:kpeakContour} we use this model to show how $\kp$ depends on $\beta$ and $\phi_i$.  Nearly all chameleon models have 
$\kp\gg 10^{10}\, \mathrm{GeV}$, and $\kp \gsim \Mpl$ for $\beta\gsim4$, casting serious doubt on the validity of the EFT 
(\ref{L}).  (This is the regime relevant for all direct detection experiments.)
Although the results shown 
in Fig.~\ref{Fig:kpeakContour} were derived assuming an exponential potential, our analytic model predicts that any chameleon potential 
$V(\phi/M)$ with $M\lsim 0.01$ eV will give similar results.
Furthermore, the values of $\kp$ in Fig.~\ref{Fig:kpeakContour} may be underestimated, because we only included contributions from SM particles in $\Sigma$.  Including additional particles, the QCD trace anomaly \cite{Kajantie:2002wa,Davoudiasl:2004gf}, interactions during the QCD phase transition \cite{Caldwell:2013mox}, or a coupling between the chameleon and a primordial magnetic field \cite{Mota:2011nh}
would increase $|\vel|$ and $\kp$.  

{\bf Conclusions--}  Cosmological dynamics in chameleon theories generically lead to a catastrophic breakdown of calculability just prior to BBN due to 
the same matter coupling that was introduced to suppress unacceptable fifth forces.  The theory can evade strong particle production effects only for weak couplings and
highly fine-tuned initial conditions, so significant advances in chameleon theory and phenomenology are required give the theory a solid 
footing.  This chameleon catastrophe is a consequence of a great mystery of modern physics: the extreme hierarchy between the masses of SM 
particles and the energy scale associated with cosmic acceleration.   We have shown how this hierarchy leads to violations of adiabaticity and the 
quantum production of particles. Other modified gravity 
theories that include scalars coupled to the trace of the stress tensor may face similar difficulties, if the effective mass of the scalar field is sensitive 
to small changes in the field's value.

\acknowledgments
{\bf Acknowledgments--} N.B. thanks DESY, the University of Geneva, and the Perimeter Institute for Theoretical Physics for their hospitality during the completion of this work.  Research at the Perimeter Institute is supported by the Government of Canada through Industry Canada and by the Province of Ontario through the Ministry of Research and Innovation.  A.E. is supported in part by the Canadian Institute for Advanced Research.  C.B. is supported by a University of Nottingham Anne McLaren Fellowship.

\end{document}